\documentclass[a4paper,pdftex,floatfix,groupedaddress,pra,showpacs,twocolumn]{revtex4}
\pdfoutput=1
\usepackage{amsfonts,amsmath,amssymb}
\usepackage{graphicx}
\usepackage{bm} 
\usepackage{mathrsfs}
\usepackage{subfigure}
\usepackage{textcomp}
\usepackage{hyperref}

\newcommand{\degree}{\ensuremath{^\circ}}%
\newcommand{\ie}{i.\,e.}%

\newlength{\figwidth}
\newlength{\figwidthlarge}
\begin{document}

\title{Ionization of 1D and 3D oriented asymmetric top molecules by intense circularly polarized femtosecond
   laser pulses}

\author{Jonas L. Hansen$^2$}%
\author{Lotte Holmegaard$^1$}%
\author{Line Kalh{\o}j$^1$}%
\author{Sofie Louise Kragh$^1$}%
\author{Henrik Stapelfeldt$^{1,2}$}%
\email[Corresponding author: ]{henriks@chem.au.dk}%
\affiliation{$^1$\,Department of Chemistry, Aarhus University, DK-8000 Aarhus C, Denmark \\
   $^2$\,Interdisciplinary Nanoscience Center (iNANO), Aarhus University, 8000 Aarhus C, Denmark}

\author{Frank Filsinger}%
\author{Gerard Meijer}%
\author{Jochen K\"upper}%
\email[Corresponding author: ]{jochen@fhi-berlin.mpg.de}%
\affiliation{Fritz-Haber-Institut der Max-Planck-Gesellschaft, Faradayweg 4-6, D-14195 Berlin,
   Germany}

\author{Darko Dimitrovski}%
\author{Mahmoud Abu-samha}%
\author{Christian Per Juul Martiny}
\author{Lars Bojer Madsen}%
\email[Corresponding author: ]{bojer@phys.au.dk}%
\affiliation{Lundbeck Foundation Theoretical Center for Quantum System Research, Department of
   Physics and Astronomy, Aarhus University, DK-8000 Aarhus C, Denmark}

\date{\today}
\begin{abstract}
We present a combined experimental and theoretical study on strong-field ionization of a three-dimensionally oriented asymmetric top molecule, benzonitrile (C$_7$H$_5$N), by circularly polarized, nonresonant femtosecond laser pulses. Prior to the interaction with the strong field, the molecules are quantum-state selected using a deflector, and 3-dimensionally (3D) aligned and oriented adiabatically using an elliptically polarized laser pulse in combination with a static electric  field. A characteristic splitting in the
molecular frame photoelectron  momentum distribution reveals the position of the nodal planes of the molecular orbitals from which ionization occurs. The experimental results are supported by a theoretical tunneling model that includes and quantifies the splitting in the momentum distribution. The focus of the present article is to understand strong-Þeld ionization 
from 3D-oriented asymmetric top molecules, in particular the suppression of electron emission in nodal planes 
of molecular orbitals. In the preceding article [Dimitrovski et al., Phys.~Rev.~A \textbf{83}, 023405 (2011)] the focus is to 
understand the strong-Þeld ionization of one-dimensionally-oriented polar molecules, in particular asymmetries 
in the emission direction of the photoelectrons.
\end{abstract}
\maketitle

\section{Introduction}

Photoelectron spectroscopy has been and remains a ubiquitous and useful method to obtain information
about the electronic structure of molecules. In the most traditional form absorption of a single photon frees a single electron
from the molecule and the kinetic energy is
recorded. The classical use of photoelectron spectroscopy is the determination of binding energies
of molecular orbitals and possibly of the vibrational substructure~\cite{Reid:ARPC:2003}. Over the past 10-15 years such spectroscopy in
combination with femtosecond pump-probe methods has been demonstrated to be also a powerful technique for obtaining time-resolved insight into the coupled electronic and
vibrational dynamics occurring after photoabsorption in polyatomic molecules \cite{stolow:2004:cr,underwood:stolow:review:2008}.

In addition to the kinetic energy, the emission direction of the electron is a highly useful
experimental observable provided it is detected for defined vibrational and rotational states of the cation formed \cite{Hockett:PRL:2009}
or if it is measured with respect to the molecular
frame~\cite{Reid:ARPC:2003, Suzuki:ARPC:2006}. One experimental approach to obtaining molecular
frame photoelectron angular distributions (MFPADs) is to determine the molecular frame after the
ionization event. This is possible if ionization leads to fragmentation of the molecule and the
electrons can be measured in coincidence with fragment ions recoiling in directions that directly
reflect the molecular orientation at the moment of ionization, a condition often referred to as the axial recoil limit. Such coincidence methods have been
applied successfully, predominantly to small molecules where synchrotron radiation removes an inner
shell electron \cite{SHIGEMASA:PRL:1995,Akoury:Science:2007,Schoffler:Science:2008,Yamazaki09}, but also to situations where valence electrons are removed by UV
radiation~\cite{Chandler-Hayden:prl:2000, Janssen-Hayden:prl:2004, stolow:2006:science}.

The other experimental approach to MFPADs is to fix the orientation prior to the ionization event,
which offers the obvious advantage that no fragmentation of the molecule is needed. Using resonant photoexcitation mild 1D alignment
can be created for molecules in excited states \cite{Suzuki:ARPC:2006,suzuki:prl:2010}. Alternatively, alignment
methods, based on the application of moderately intense, nonresonant laser pulses, can confine the principal
axes of polarizability of molecules sharply along space-fixed axes
\cite{stapelfeldt:2003:rmp} and, thereby, provide the desired fixed-in-space targets for
photoionization of molecules possessing an inversion center. The first experiments showing MFPADs measurement from
aligned, inversion symmetric (unpolar) linear molecules were reported 2008-2009
\cite{meckel:Science:2008,Kumarappan:PRL:2008,bisgaard:Science:2009}. Targets of molecules lacking an inversion center (i.e., polar molecules)
must, in addition to having their axes confined, possess a directional order of their dipole moment,
\ie, the molecules must be oriented as well as aligned \cite{stapelfeldt:2003:rmp}. One approach
to efficient orientation and alignment is through the combined
action of a laser pulse and a weak static electric field \cite{friedrich:1999:jcp,sakai:2003:prl,ghafur:natphys:2009,Holmegaard:PRL102:023001}.

In the present work we study single ionization of 1D and 3D aligned and oriented samples of the prototypical asymmetric top molecule benzonitrile by intense, circularly polarized femtosecond laser pulses, extending  our first results reported recently \cite{Holmegaard:NatPhys:2010}. For 1D aligned and oriented molecules pronounced up-down asymmetries of the electron emission are observed resulting from a higher probability for ionization when the circularly polarized field points along the permament dipole moment of benzonitrile compared to the opposite direction. These results are similar to those obtained for the linear OCS molecules reported in the accompanying paper \cite{OCS_circ}. For 3D aligned and oriented molecules striking suppression of electron emission in the polarization plane of the ionizing laser is observed. Our theoretical analysis, relying on a modified tunnel ionization theory,  identifies the suppression as resulting from nodal planes in the orbitals from which electrons are removed. Note that effects of nodal plane structure and orbitals below the highest occupied molecular orbital on strong field ionization of in symmetric molecules was discussed theoretically some time ago~\cite{kjeldsen:2003:pra,kjeldsen:2005:pra}.

The paper is organized as follows. In Sec.~II, the experimental setup is discussed. The experimental results are presented in Sec.~III, including a discussion of the method of deflection, alignment and orientation of the benzonitrile molecules, and the MFPADs from strong-field single ionization of benzonitrile by a circularly polarized femtosecond laser field. Section IV contains the theory including a discussion of the characteristic off-nodal-plane emission pattern. In Sec.~V, the theoretical MFPADs  are compared with the experimental ones. Section VI concludes and gives a brief outlook. The molecular properties of benzonitrile, used in the theoretical model, are summarized in Appendix A.

\section{Experimental setup}
\label{setup}

\subsection{Molecular beam apparatus}
\label{Molecular-beam-apparatus}

\begin{figure}
   \centering
   \includegraphics[width=\figwidth]{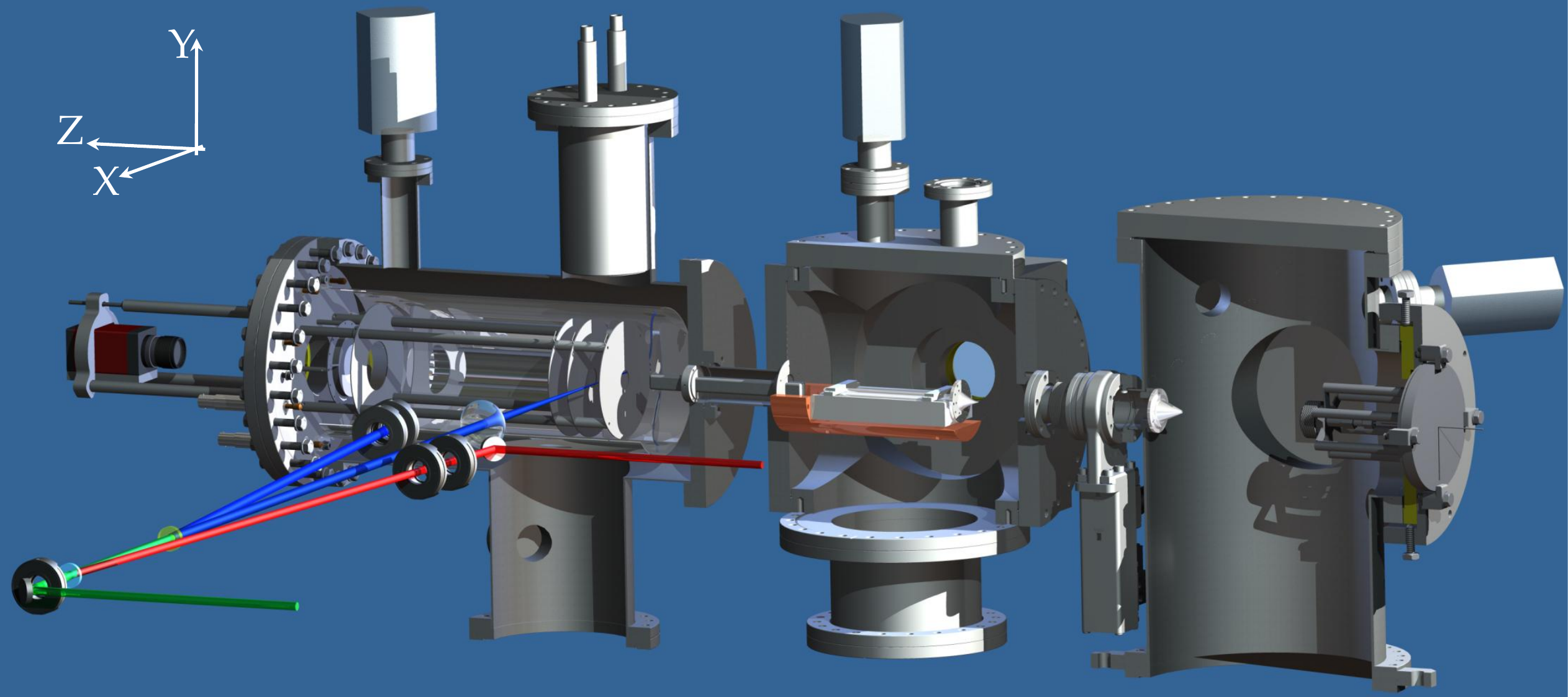}
   \caption{(Color online) Exploded view of the molecular beam machine consisting of (from the
      right) the source chamber, the deflector chamber and the target chamber. In this work two
      laser beams are used, the 1064 nm YAG beam (blue cylinder) and the 800~nm probe beam (red
      cylinder). An additional 800~nm femtosecond beam (green cylinder) can be included to induce,
      for instance, rotational dynamics. The pair of circular discs in each laser beam represents a
      half-wave and a quarter-wave plate used to control the
      polarization state of the laser pulses.}
   \label{fig:vacuum-chamber}
\end{figure}

An exploded schematic of the molecular beam apparatus, used in these studies, is shown in
\autoref{fig:vacuum-chamber}.
It consists of three vacuum chambers: the source chamber, pumped by a 2000 l/s turbomolecular pump,
followed by the deflector and the target chamber which are both pumped by 500 l/s turbomolecular
pumps. The molecular beam is formed by expanding a mixture of carrier gas (typically 90 bar of
Helium) and the molecular gas ($\sim$5 mbar of benzonitrile) into the source chamber through a
high-pressure Even Lavie valve \cite{even:2000:jcp,even:2003:jcp} heated to 40\,\degree{C}. Operating the valve at high
stagnation pressure and fine tuning the valve opening time ensures formation of a cold molecular
beam without cluster formation. We estimate the rotational temperature of the molecular gas with two
independent techniques. One relies on comparing experimentally observed degrees of alignment with
theoretical calculations for a temperature averaged sample of molecules \cite{kumarappan:2006:jcp},
whereas in the other method vertical molecular beam profiles of electrostatically deflected
molecules are measured and compared with simulations \cite{Filsinger:JCP:2009}. We find that the
typical rotational temperature is $\sim$1~K. The experiments are performed at 20~Hz limited by the repetition rate of the
alignment laser. The molecular beam is skimmed
twice, 14 and 38 cm downstream from the nozzle of the Even-Lavie valve, respectively, to collimate the expansion. The
first skimmer, positioned at the exit of the source chamber, has a 3-mm diameter (Beam Dynamics
model 50.8). In the deflector chamber the molecular beam passes through the second skimmer
(1-mm diameter Beam Dynamics model 2) just before entering the 15-cm-long electrostatic deflector.

\subsection{Electrostatic deflector}
\label{deflector}

The deflector consists of a trough-shaped electrode with a 3.2-mm-inner-radius of curvature kept at
ground potential and a polished rod with a radius of 3.0 mm to which high voltage (up to 10 kV) can
be applied \cite{Filsinger:JCP:2009}. The rod is kept floating above the trough by two pieces of macor, providing a minimum
distance of 0.9 mm between the electrodes and 1.4-mm-separation at the molecular beam axis. This
electrode geometry creates a two wire field, that is strongly inhomogeneous along the (vertical) Y-axis
while it is almost homogeneous in the X direction~\cite{ramsey:molecular_beams}. Therefore, polar molecules traversing
the deflector experience a force in the vertical direction with molecules
in high-field seeking states being deflected upwards. The extent of deflection for a molecule depends on
the effective dipole moment of the particular molecular quantum state.
As the molecules exit the deflector, they enter a parallel plate
capacitor consisting of two 17-cm-long polished stainless steel plates separated by 2.7~mm. The
plates are oriented parallel to the deflection direction. The capacitor creates a constant dc
electric field on the order of 2~kV/cm along the path of the molecules moving from the deflector to
the laser interaction zone. This finite field prohibits population transfer between quantum states
due to Majorana transitions between $M$-levels of individual rotational states at zero field, and it
mitigates transfer due to diabatic traversal of avoided crossings induced by rotations or
inhomogeneities of the electric field, which occur frequently at small field
strengths \cite{Wohlfart:PRA78:033421,Kirste:PRA:2009,Wall:PRA:2010}

\subsection{ Velocity map imaging spectrometer and laser beams}
\label{deflector}

The target chamber houses a velocity map imaging (VMI) spectrometer consisting of an open
three-electrode electrostatic lens \cite{eppink:1997}. Between the first two electrodes (repeller
and extractor) the molecular beam is crossed by two (or more) pulsed laser beams as indicated with
the red and blue (and green) cylinders in \autoref{fig:vacuum-chamber}. Applying positive (negative)
voltages to the electrostatic lens ions (electrons) in the interaction region are accelerated
towards a 50-mm-diameter microchannel plate (MCP) backed by a similar size phosphor screen. The ion
(or electron) images are recorded by a CCD camera monitoring the fluorescent phosphor screen, and
on-line software analysis determines and saves the coordinates of each individual particle hit. When
imaging ions the MCP detector can be gated by applying a fast high-voltage pulse to its front
thereby obtaining images for a selected range of mass-to-charge ratios. The spectrometer is caged in a single concentric
mu-metal cylinder to minimize the influence of external magnetic fields on the trajectories of the
electrons.

The laser systems, as well as the spatial and temporal overlap of different laser beams with the
molecular beam is similar to that reported before \cite{kumarappan:2006:jcp} so the description here
is brief. The fundamental output (1064 nm) from a 20 Hz, Nd:YAG laser is used to adiabatically align
the molecules. The pulse duration is 10 ns. Part of the output from a 1 kHz amplified Ti:sapphire
laser (800 nm) is compressed to 30 fs and used to either Coulomb explode the molecules
\autoref{alignment-orientation}, for detecting their spatial orientation or to singly ionize the
molecules for creating photoelectrons (\autoref{sec:PAD-BN}) or for characterizing the molecular beam
profile (\autoref{deflection}). The alignment beam (hereafter termed the YAG beam or the YAG pulse)
and the fs beam (hereafter termed the probe beam or the probe pulse) are collinearly overlapped
using a zero-degree mirror (high reflectivity at 1064 nm and high transmission at 800 nm) and
focused onto the molecular beam by a 50~mm-diameter 30~cm-focal-length spherical lens. The probe
pulse is electronically synchronized to the peak of the YAG pulse.

Using telescopes in each laser beam the spotsizes are adjusted so that the probe beam is more
tightly focused than the YAG beam ($\omega_0 ^\text{YAG} = 34~\mu$m and $\omega_0 ^\text{probe} = 21~\mu$m,
measured by scanning a 10 $\mu$m pinhole). The intensity of
the probe pulse is adjusted by inserting thin neutral density filters in the beam, the intensity of
the YAG pulse by rotating a half wave plate placed between two thin-film
polarizers, upstream, and one thin film polarizer downstream. The polarization state of either
beam is controlled by a half-wave and a quarter-wave plate inserted in each beam just before the
final 0 degree overlap mirror.

\section{Experimental results}

\subsection{Deflection of benzonitrile}
\label{deflection}

\begin{figure}
   \centering
   \includegraphics[width=\figwidth]{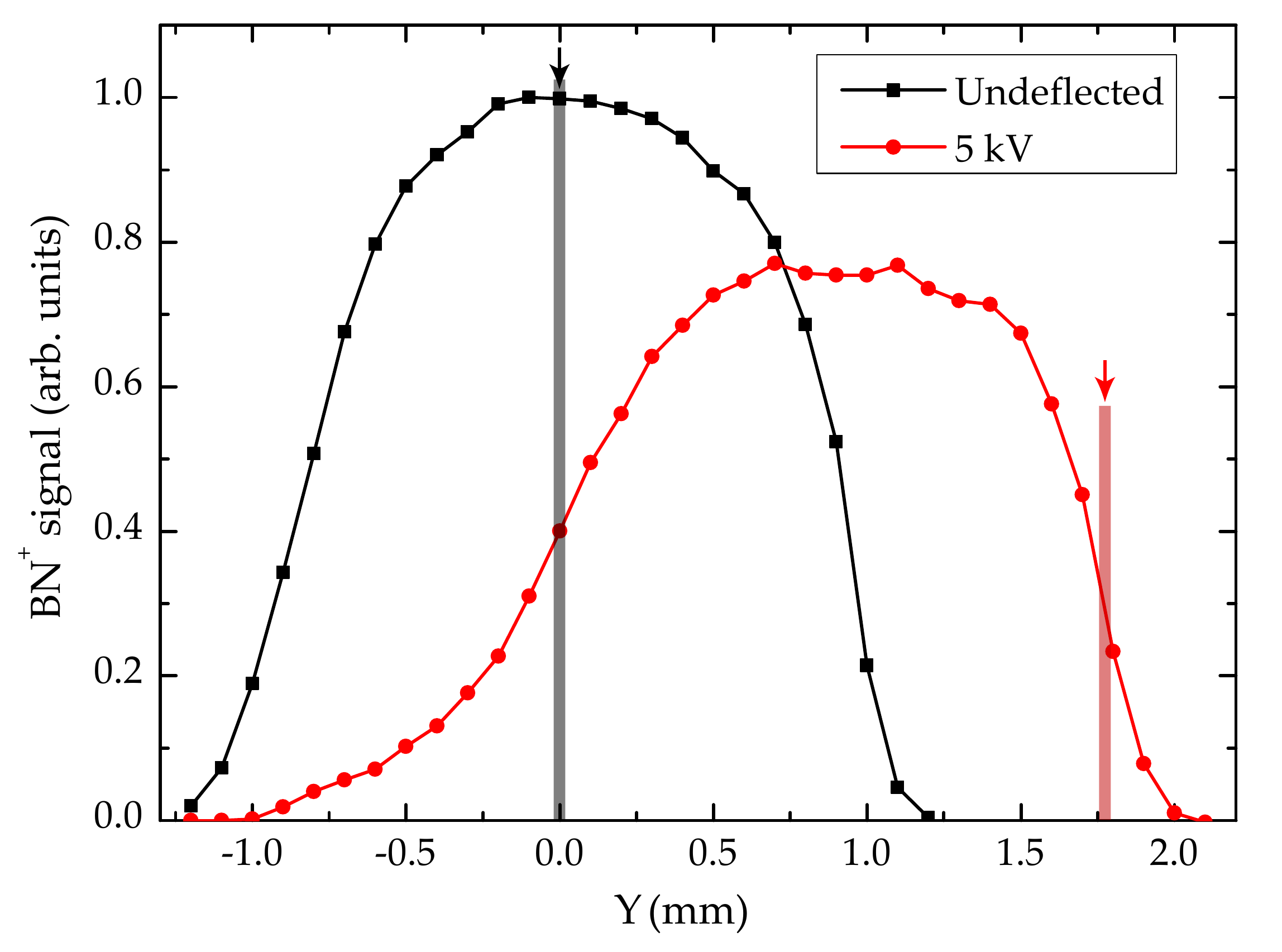}
   \caption{(Color online) Vertical profiles of the molecular beam measured by recording the
      BN$^{+}$ signal as a function of the vertical position of the fs probe beam focus. The
      experimental data are shown by black squares (deflector off, 0~kV) and red circles (5~kV).
      The connecting lines serve to guide the eye. The
      arrows indicate the position of the laser foci for acquiring ion and electron images of
      undeflected (black), Y~=~0~mm, and deflected (red), Y~=~1.55~mm, molecules.
      }
   \label{fig:deflection}
\end{figure}

The effect of the deflector on the molecular beam is illustrated by the vertical intensity profiles
in \autoref{fig:deflection}. These profiles are obtained by recording the BN$^+$ signal [BN = benzonitrile (C$_7$H$_5$N)] from
photoionization due to the femtosecond probe laser ($I_\text{probe}\approx10^{14}$~W/cm$^2$) as a
function of the vertical position of the laser focus. When the deflector is turned off, the
molecular beam extends over $\sim1.5$~mm, mainly determined by the diameter of the skimmer before
the deflector and the end-aperture of the deflector. When the deflector is turned on, the
molecular beam profile broadens and shifts upwards (positive Y values). Molecules in the lowest
rotational quantum states have the largest Stark shifts and are, therefore, deflected the
most~\cite{Holmegaard:PRL102:023001, Filsinger:JCP:2009}. In the measurements described below,
experiments were conducted on these quantum-state selected molecules simply by positioning the laser
foci close to the upper cut-off region in the 5~kV profile (Y~=~1.55~mm), as indicated by the red
arrow in \autoref{fig:deflection}.

\subsection{Alignment and orientation}
\label{alignment-orientation}

The molecules studied here are asymmetric tops. Their alignment and orientation is controlled by the
combination of the YAG pulse and the weak static electric field in the VMI spectrometer. Based on
previous experiments and theory a linearly polarized YAG pulse is used to induce 1D alignment and
orientation whereas 3D alignment and orientation is created by applying an elliptically polarized
YAG pulse \cite{larsen:1999:jcp, larsen:2000:prl, stapelfeldt:2003:rmp, tanji:pra:2005, Holmegaard:PRL102:023001,
   Nevo:PCCP:2009}. Here 1D alignment refers to confinement of a single molecular axis along the YAG
polarization axis. Due to the nature of the polarizability
interaction between the molecule and the linearly polarized
alignment field it is the most polarizable axis that is aligned
\cite{stapelfeldt:2003:rmp}. For benzonitrile, discussed here,
this is the C$_2$-axis (\ie, the C-N axis). Similarly, 3D
alignment refers to confinement of three perpendicular molecular
axes \cite{stapelfeldt:2003:rmp}. In practice, the elliptically
polarized alignment field confines the most polarizable axis along
the major polarization axis, the second most polarizable axis (for
BN the axis in the molecular plane perpendicular to the
C$_2$-axis) along the minor polarization axis and, consequently,
the least polarizable axis (for BN the axis perpendicular to the
molecular plane) perpendicular to the polarization plane. Finally,
orientation refers to whether the dipole moment of the molecules
point in a particular direction with respect to a laboratory-fixed
reference direction, which here is the weak static electric field
in the VMI spectrometer.

To test that alignment and orientation of BN occurs, the molecules are Coulomb exploded by the
intense fs probe pulse ($I_{\text{probe}}~=~5.4\times10^{14}$~W/cm$^2$) and recoiling ionic fragments
are recorded by the 2D ion detector, a method used previously for many other molecules in our
laboratory. For a planar molecule of C$_\text{2v}$ symmetry, like BN, our standard procedure is to
first establish that the C$_2$-axis is aligned. This is accomplished by recording CN$^+$ ions
because they primarily recoil along the C$_2$-axis.
\begin{figure}
    \centering
   \includegraphics[width=\figwidth]{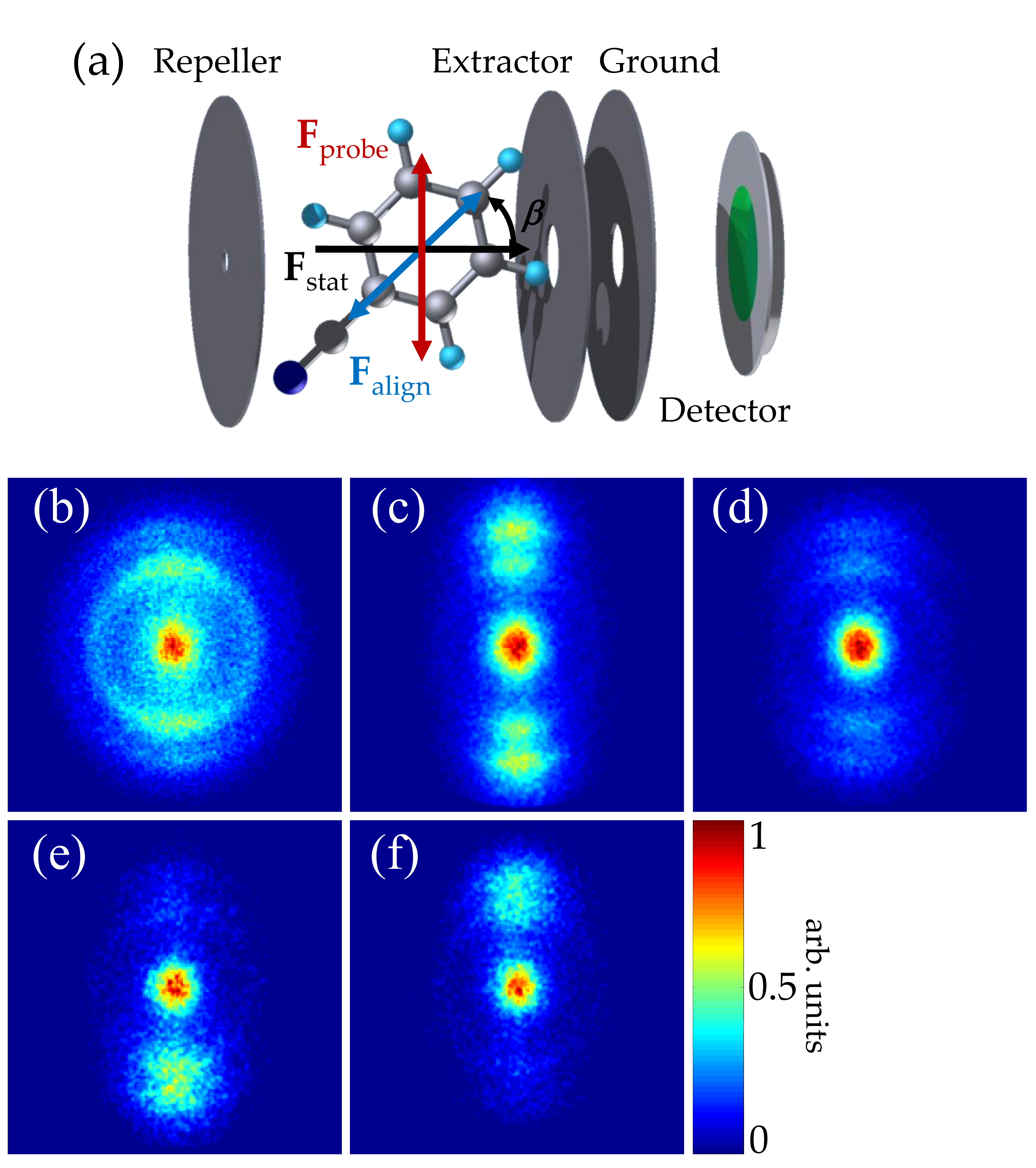}
   \caption{(Color online) (a) Schematic of the velocity map imaging spectrometer used to detect ions or
      electrons. The 1D alignment of the BN molecules is determined by the alignment laser
      polarization, here shown for $\beta=45^\circ$. The static electric field of the spectrometer,
      pointing from the repeller to the extractor electrode for ion detection, breaks the
      head-for-tail symmetry by preferentially placing the CN-end towards the repeller where the
      positive voltage is highest. When detecting electrons the polarity of the electrodes is
      inverted forcing the CN-end of the molecules towards the extractor electrode.
      (b)-(f) CN$^+$ ion images. (b): probe pulse only, vertically polarized.
      (c), (d) $\beta=90^\circ$. In (d) the rotational temperature is
      significantly higher than in (c) and no quantum state selection occurs (see text).
      (e), (f): $\beta=45^\circ$ and -45$^\circ$, respectively. For all images
      $\text{F}_{\text{stat}}=286$ V/cm . In (e) / (f) 71~\% / 28~\% of all CN$^+$
      ions appear in the upper half of the detector. }
  \label{BN-alignment-orientation}
\end{figure}
\autoref{BN-alignment-orientation}(b) shows a CN$^+$ image
recorded with the probe pulse only, linearly polarized in the
detector plane. A mild confinement of the ions is observed
reflecting the enhanced ionization probability for molecules that
initially have their C$_2$-axis pointing along the probe pulse
polarization. The confinement is quantified by
$\langle\cos^2\theta_\text{2D}\rangle$~=~0.63, where
$\theta_\text{2D}$ is the angle between the projection of the
CN$^+$ recoil velocity on the detector plane and the YAG
polarization~\cite{kumarappan:2006:jcp}. When the YAG pulse is
included, linearly polarized vertically, corresponding to
$\beta=90^\circ$ ($\beta$ denoting the angle between the YAG
polarization and the static electric field), the angular
confinement sharpens markedly
[\autoref{BN-alignment-orientation}(c)] and
$\langle\cos^2\theta_\text{2D}\rangle$ increases to 0.89. This
shows that the BN molecules are 1D aligned with their C$_2$-axis
along the YAG polarization. The
$\langle\cos^2\theta_\text{2D}\rangle$ value is quite a bit
smaller than that observed for 1D alignment of iodobenzene
($\langle\cos^2\theta_\text{2D}\rangle=0.97$) \cite{Holmegaard:PRL102:023001} under similar
conditions of initial rotational cooling, quantum-state selection
and alignment pulse intensity. This may seem surprising since the
components of the polarizability tensor for BN are similar to
those of iodobenzene and thus the two molecules should exhibit
approximately the same degree of 1D adiabatic alignment. The
reason for the lower $\langle\cos^2\theta_\text{2D}\rangle$ value
for BN is most likely that CN$^+$ ions are not ideal observables
for deducing the spatial orientation of the C$_2$-axis because,
unlike I$^+$ ions used in the iodobenzene characterization, it
overlaps with other fragment ions in the time of flight spectrum
of C$_7$H$_5$N. Hence, the ion images of
\autoref{BN-alignment-orientation} also contain contributions from
hydrocarbon fragments from the benzene ring of equal (or nearly
equal) mass-to-charge-ratio (e.g. C$_2$H$_2^+$). These
``contaminant'' ions are not expected to recoil along the C-CN
axis. In the alignment measurements we try to minimize the
contribution from the contaminant ions by only analyzing the
CN$^+$ ions with the highest kinetic energy corresponding to the
outermost ring~\cite{footnote1} seen in the images of
\autoref{BN-alignment-orientation}. However, just a small
contributions from the contaminant ions ejected in directions away
from the YAG polarization is enough to significantly lower the
observed $\langle\cos^2\theta_\text{2D}\rangle$ value.

For reasons discussed in Sec. \ref{sec:PAD-BN} alignment was also
measured for conditions of higher rotational temperature and no
rotational state selection. In practice, this is achieved by
lowering the He stagnation pressure from 85 to 15 bar
\cite{kumarappan:2006:jcp} and, in addition, turning off the
deflector voltages. The pronounced decrease in the degree of alignment is
clear by comparing Fig. \ref{BN-alignment-orientation}(d) (15 bar
He, deflector turned off) with Fig. \ref{BN-alignment-orientation}
(c) (90 bar He, selection of lowest rotational quantum states).
Quantitatively, $\langle\cos^2\theta_\text{2D}\rangle$ is 0.76 and
0.89 in Figs. \ref{BN-alignment-orientation} (d) and (c),
respectively.

\begin{figure}
    \centering
   \includegraphics[width=\figwidth]{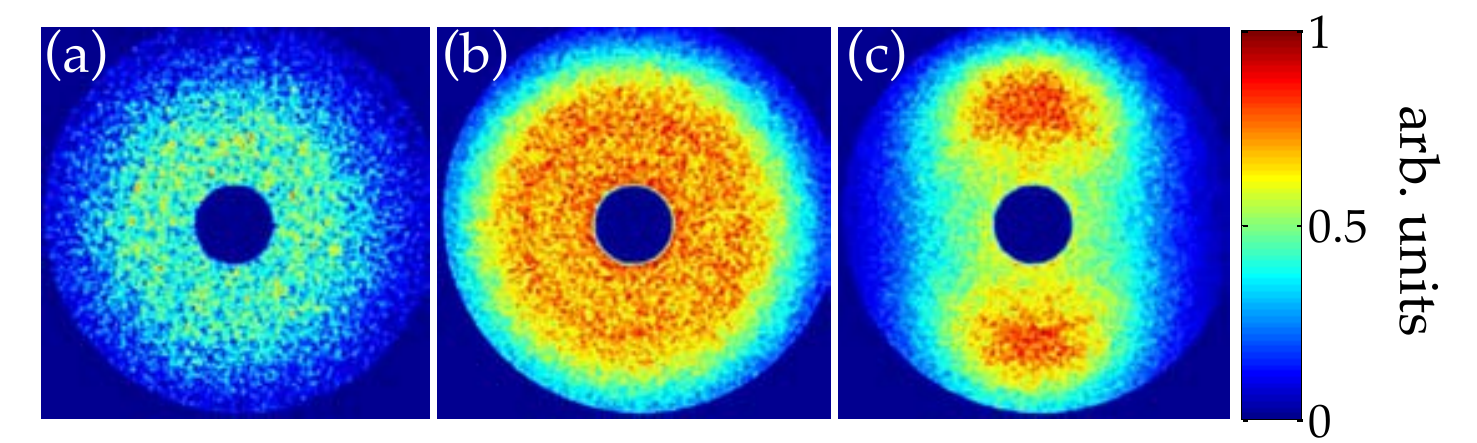}
   \caption{Ion images of H$^+$ recorded with the probe laser ($5.4\times10^{14}$~W/cm$^2$) linearly polarized perpendicular
to the detector plane. (a) No YAG pulse. (b) YAG pulse included,
linearly polarized perpendicular to the detector. (c) YAG pulse
included, elliptically polarized. The major axis is perpendicular
to the detector and the ellipticity-intensity ratio is 3:1. The
center has been removed to suppress a large spike of very low
kinetic energy H$^+$ ions stemming from ionization of residual
molecules in the target chamber.
}
  \label{BN-3D-H+}
\end{figure}

To achieve 3D alignment the polarization of the YAG pulse is
changed from linear to elliptical. When an ellipticity of 3:1 is
chosen, the value referring to the ratio of the intensities
measured along the major and the minor polarization axis, the
CN$^+$ ion image (not shown) recorded with the major polarization
axis in the plane of the detector (vertical) is very similar to
the image in \autoref{BN-alignment-orientation} (c), showing an
equally strong degree of 1D alignment of the C$_2$-axis. To
additionally confirm confinement of the molecular plane H$^+$ ions
were recorded, see \autoref{BN-3D-H+}, following an approach
employed in previous studies
\cite{Viftrup:prl:2007,Nevo:PCCP:2009} of 3D alignment. When the
probe laser only is used the image is circularly symmetric [Fig.
\ref{BN-3D-H+} (a)]. Including the linearly polarized YAG pulse,
polarized perpendicular to the detector plane, does not change the
circular symmetry [Fig. \ref{BN-3D-H+} (b)]. This is consistent
with the C$_2$-axis being aligned along the YAG polarization
vector and the molecular plane free to rotate around the arrested
axis. When the 3:1 elliptically polarized YAG pulse, its major
polarizability axis perpendicular to the detector, is used, the
circular symmetry is broken and, instead, the H$^+$ ions localize
around the vertical axis corresponding to the minor axis of the
polarization ellipse. Similar to previous studies \cite{Viftrup:prl:2007,Nevo:PCCP:2009} we interpret
this as confinement of the molecular plane with the second most
polarizable axis (perpendicular to the C$_2$-axis but still in the
molecular plane) aligned along the minor axis of the ellipse.

To show orientation of BN the polarization of the linearly
polarized YAG pulse is rotated to $\beta=45^\circ~(-45^\circ)$
meaning that the C$_2$-axis of the molecules is positioned at 45
(-45) with respect to the static electric field,
$\textbf{F}_{\text{stat}}$, of the VMI spectrometer [see Fig.
\ref{BN-alignment-orientation}(a)].
The resulting CN$^+$ images
are displayed in Figs. \ref{BN-alignment-orientation}(e) and (f).
The appearance of more CN$^+$ ions on the lower (upper)
part of the detector for 45 (-45) shows that the molecules are
oriented with the CN-end preferentially pointing towards the
repeller electrode. Hence, the permanent dipole moment of BN is
pointing in the direction from the repeller to the extractor as
expected. The up-down asymmetry remains unchanged when an
elliptically rather than a linearly polarized YAG pulse is
employed to align the molecules. These findings of alignment and orientation
are fully consistent with recent studies on
iodobenzene \cite{Holmegaard:PRL102:023001,Filsinger:JCP:2009} and
2,6-difluoroiodobenzene \cite{Nevo:PCCP:2009}.

The degree of orientation is characterized by the ratio of ions detected on the upper part to the
total number of ions detected. For $\beta=45^\circ$ it is 27~\% and for
$\beta=-45^\circ$ it is 71~\%. For several reasons these numbers underestimate the degree of orientation utilized in the
photoelectron experiments described below. The presence of contaminant ions with the same
mass-to-charge ratio as CN$^+$ will reduce the up-down asymmetry. In addition, the ion images are
recorded at F$_\text{stat}$=286~V/cm which is significantly lower than the electrostatic extraction
field in the photoelectron angular distribution (PAD) measurements presented in \autoref{sec:PAD-BN}
(F$_\text{stat}=467$~V/cm). In fact, the increase of the static field will be even larger because
the ion measurements are recorded at $\beta=45^\circ$ , \ie, the effective value of
\textbf{F$_\text{stat}$} is reduced by $\sqrt{2}$. By contrast for the PAD measurements the
alignment field is polarized parallel to \textbf{F$_\text{stat}$}. As a consequence, we estimate
that the orientation in the PAD geometry corresponds to at least 80 \% of the molecules have their
CN-end facing the extractor electrode.

\subsection{PADs from single ionization of benzonitrile}
\label{sec:PAD-BN}

For the PAD experiments the same experimental setup as for the ion detection described in
\autoref{alignment-orientation} is used, but some essential parameters are changed. The polarization
state of the 30 fs probe pulses is changed from linear to circular and the intensity is lowered to
$\simeq1.2\times10^{14}$~W/cm$^2$ corresponding to a regime where the BN molecules only undergo single
ionization with essentially no fragmentation. Also, the polarization of the alignment pulse is
changed such that the major axis is parallel to the static field axis. Furthermore, to extract
electrons instead of ions in the PAD measurements the polarity of the velocity map imaging
spectrometer is inverted. Hereby, the BN molecules are confined along the static field axis with the
CN-end facing the extractor electrode.

\begin{figure*}
    \centering
   \includegraphics[width=\figwidthlarge]{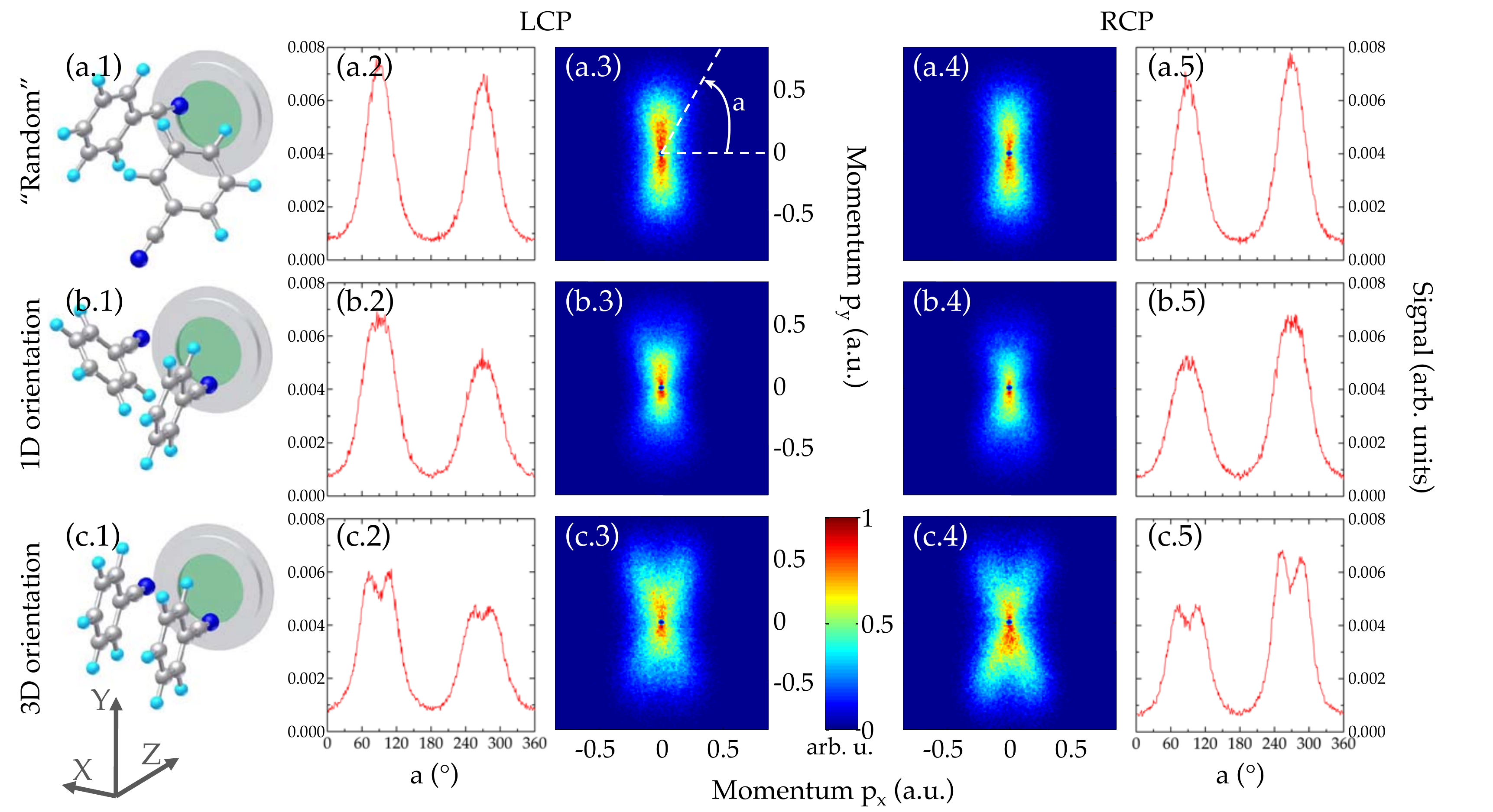}
   \caption{Two-dimensional momentum images of electrons produced when BN molecules are ionized by a
      LCP probe pulse or a RCP probe pulse. The corresponding angular distributions in the detector
      plane are shown next to the images. In row (a) the molecules are essentially randomly oriented
      (no alignment pulse). In row (b) the molecules are 1D aligned and oriented (linearly polarized
      alignment pulse) and in row (c) the molecules are 3D aligned and oriented (elliptically
      polarized alignment pulse). The intensity of the probe pulse is $1.2\times10^{14}$~W/cm$^2$.
      The intensity of the YAG pulse is $7\times10^{11}$~W/cm$^2$. In row (c) the intensity ratio
      between the major and minor axis is 3:1. The (X,Y,Z) coordinate system is oriented identical to the one shown in Fig. \ref{fig:vacuum-chamber}.
      }
  \label{fig:BN-PADS}
\end{figure*}

The photoelectron images are shown in \autoref{fig:BN-PADS}. Applying only the probe pulse [Figs. \ref{fig:BN-PADS}(a.3)
and (a.4)] the electrons emerge in a stripe parallel to the polarization plane (Y,Z) of the probe pulse
for both left and right circularly polarized (LCP and RCP) pulses. A weak up-down
asymmetry is observed, that reverses as the helicity of the
probe pulse is flipped. The asymmetry is more easily seen in the angular distributions
[Figs. \ref{fig:BN-PADS}(a.2) and (a.5)], obtained by radially integrating the respective images from momentum 0.02 to 0.8
a.u. When the YAG pulse, linearly polarized along {\bf F$_\text{stat}$}, is
included to induce strong 1D alignment and orientation, schematically illustrated in
\autoref{fig:BN-PADS}(b.1), the up-down asymmetry increases [Figs. \ref{fig:BN-PADS}(b.3) and (b.4)]. For LCP (RCP) probe
pulses the number of electrons detected in the upper part compared to the total number in the image
is $\sim$55\% (44\%). The enhanced asymmetry is also clearly visible in the angular distributions
[Figs. \ref{fig:BN-PADS}(b.2) and (b.3)]. As explained in \autoref{num-results} the up-down asymmetry results from the orientational dependence of the
ionization yield and the force on the electrons, by the probe laser field, following their
detachment from the molecule. In particular, we show that ionization of molecules, 1D aligned and
oriented with their dipole moment along the direction of {\bf F$_\text{stat}$}, leads to electrons
with an upward (downward) momentum for LCP (RCP) probe pulses in agreement with the observations.
The observation of an asymmetry without the YAG pulse results from mild orientation
due to the interaction between the permanent dipole moment and the weak static field and/or state selection by the deflector
\cite{Nevo:PCCP:2009}.

\begin{figure*}
    \centering
   \includegraphics[width=120mm]{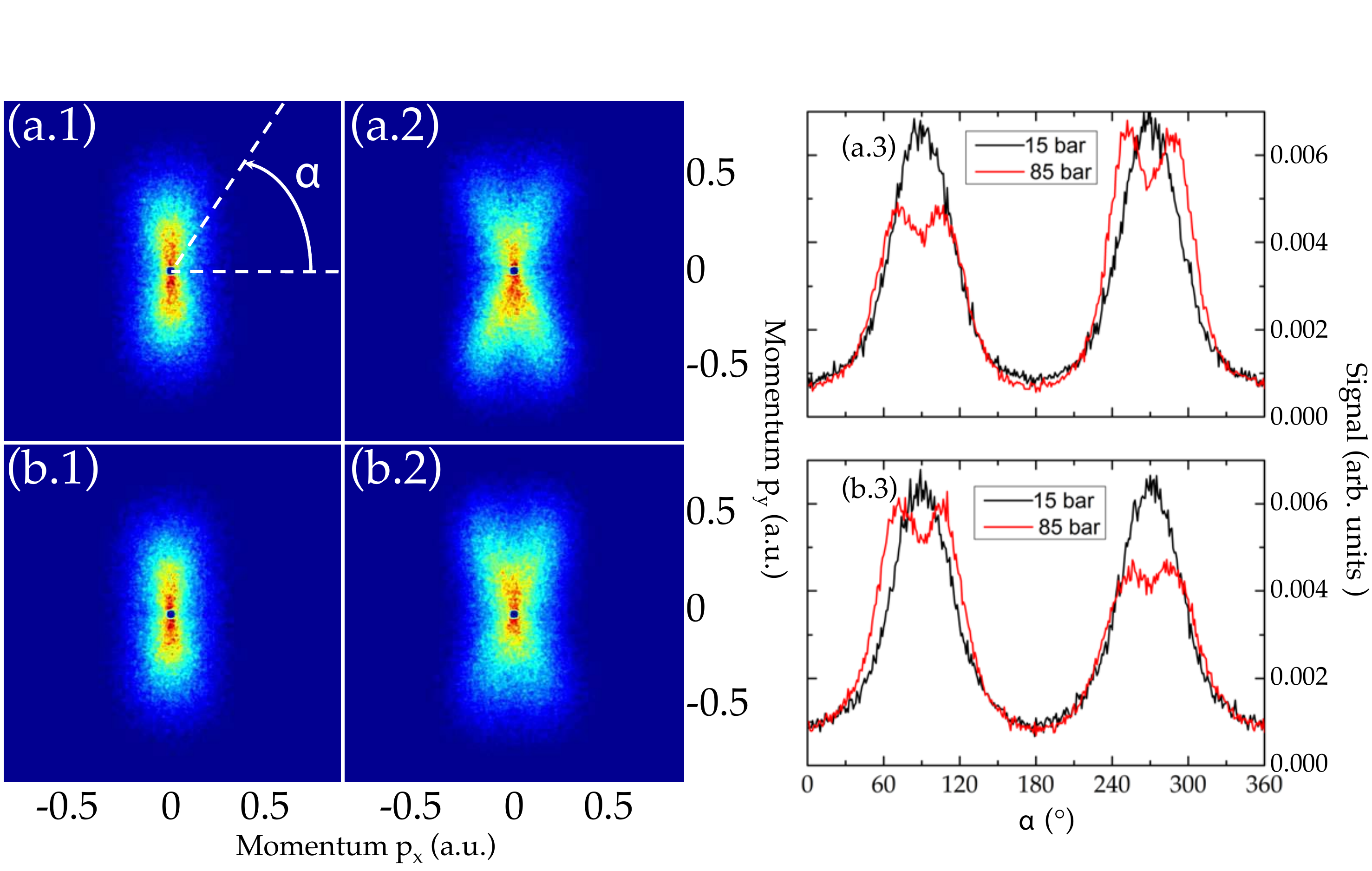}
   \caption{Two-dimensional momentum images of electrons produced when BN molecules are ionized by a
      RCP probe pulse (a.1), (a.2) or a LCP probe pulse (b.1), (b.2). The images in column 1 (2) are
      recorded for weakly (strongly) 3D aligned and oriented molecules obtained by using 15 (85) bar
      of He stagnation pressure (see text). The laser pulse parameters were the same as those used
      in Fig. \ref{fig:BN-PADS}. The angular distributions in the detector plane are compared in (a.3) and (b.3).
      }
  \label{fig:bad-good-cooling}
\end{figure*}

When the BN molecules are 3D aligned and oriented, obtained by using an elliptically rather than a
linearly polarized YAG pulse (see Sec. \ref{alignment-orientation}), striking substructures appear
in the PADs [Figs. \ref{fig:BN-PADS}(c.3) and (c.4)]. Notably, the electron emission is suppressed along the vertical
direction, clearly seen as a dip in the corresponding angular distributions [Figs. \ref{fig:BN-PADS}(c.2) and (c.5)] at
$\alpha~=90^\circ$ and $270^\circ$, where $\alpha$ is the angle measured counterclockwise with respect to the
X direction [see Fig. \ref{fig:BN-PADS}(a.3)]. The angular distributions also show that the maximum in the electron
emission occurs at an angle
\begin{equation}
\label{omega_exp}
\Omega_\text{exp} \sim 18^\circ
\end{equation}
 away from vertical. As explained in \autoref{num-results} the local minimum of electrons along the vertical
direction results from suppression of electron emission in the nodal plane of the highest-occupied molecular orbital (HOMO) [and the first orbital below HOMO in energy, HOMO-1] which, for the 3D aligned and oriented molecules, corresponds to the (Y,Z) plane.
Regarding the up-down asymmetry of the electron distributions [panels (c.2) and (c.5)] it is essentially the same as for
the electron distributions obtained for the 1D aligned and oriented molecules [Figs. (b.2) and (b.5)].
This is in accordance with expectations since the degree of 1D alignment and orientation should be
essentially the same for the linearly and the elliptically polarized YAG pulse (verified experimentally).

To test that the observed structures in the PADs are not due to perturbation from the YAG pulse
on the ionization process by the fs probe pulse, experiments were carried out on molecules with a
much lower degree of alignment but identical YAG pulse intensity. This is achieved by employing
molecules with higher rotational temperature, achieved as explained in Sec. \ref{alignment-orientation}.
Figure \ref{fig:bad-good-cooling} shows the photoelectron images obtained for LCP and RCP when the
molecules are 3D aligned and oriented. Images similar to those in Fig. \ref{fig:BN-PADS} are kept as a
reference [Figs. \ref{fig:bad-good-cooling}(a.2) and (b.2)]. Both the up-down asymmetry and the lobe-like structures are completely
absent in Figs. \ref{fig:bad-good-cooling}(a.1) and (b.1), thereby showing the importance of a tightly aligned and oriented target
for observation of the effects in the PADs. The difference between the experiments on
strongly versus weakly aligned and oriented molecules is also obvious from the angular distribution
displayed in Figs. \ref{fig:bad-good-cooling}(a.3) and (b.3).

\section{Theory}
\label{num-results}

\subsection{Stark-shift corrected tunneling theory}
\label{Stark}

To describe the response of molecules with large polarizabilities and dipole moments to strong, near infrared laser pulses requires the Stark shifts of the energy levels be included, see Ref.~\citealp{OCS_circ} and the Supplementary Information in  Ref.~\citealp{Holmegaard:NatPhys:2010}. In Ref.~\citealp{OCS_circ}, we  modified the tunneling model to include the Stark shifts of the molecule and its cation at the instantaneous value of the external field. We briefly summarize the model here  for completeness.

When the Stark shifts are taken into account,
the field-free ionization potential $I_p (0)$ is modified  and  reads (up to second order in the
field strength $\mathbf{F}$)
\begin{equation}
I_p (\mathbf{F}) = I_p (0) +{\boldsymbol \Delta \boldsymbol \mu} \cdot
\mathbf{F} +\frac{1}{2} \mathbf{F}^\text{T} {\boldsymbol \Delta} {\boldsymbol \alpha}
{\mathbf{F}} , \label{eq:2}
\end{equation}
where
\begin{equation}
{\boldsymbol \Delta} {\boldsymbol \mu} =
{\boldsymbol \mu}^{M} - {\boldsymbol \mu}^{I}  \quad
{\boldsymbol \Delta} {\boldsymbol \alpha} = {\boldsymbol
\alpha}^{M} - {\boldsymbol \alpha}^{I} , \label{eq:3}
\end{equation}
${\boldsymbol \mu}$ is the permanent dipole moment, ${\boldsymbol \alpha}$ is the polarizability
tensor, and the superscripts $M$ and $I$ are referring to the molecule and the unrelaxed cation, respectively. We use the unrelaxed cation, since the nuclei have no time to move to the new equlibrium position during the tunneling process.
At the field strengths of concern the shifts induced by the hyperpolarizability and higher-order terms are negligible compared to the linear and quadratic terms accounted for in Eq.~\eqref{eq:2}. We note that the dipole moment and polarizability differences in Eq.~\eqref{eq:3} are exactly the dipole and polarizability of the orbital actively involved in the tunneling process, e.g., the HOMO and HOMO-1 in the present case. Hence the discussion could equivalently be performed referring to the orbital dipole moment and polarizability. Such an approach is natural in the case of calculations with the time-dependent Schr{\"o}dinger equation within the single-active electron approximation~\cite{Mahmoud2010}.

In the tunneling regime, \ie, when the Keldysh parameter $\gamma$ \cite{Keldysh} is
smaller than unity, the static Stark shift should be accounted for and the standard tunneling model of Ref.~\cite{Ammosov:1986:JETP}  modified accordingly by using the shifted ionization potential of Eq.~\eqref{eq:2}. These changes are crucial in order to reproduce the experimental asymmetry in the photoelectron momentum distribution~\cite{Holmegaard:NatPhys:2010}.
First, for simplicity we restrict for simplicity our consideration to the case were the target molecule can be modeled by an s-orbital. We include a
factor to account for the saturation of ionization at  high intensities~\cite{Tong:2005:JPB}, and
the Stark-corrected tunnel ionization rate (up to a multiplicative constant) becomes
\begin{align}
w( \mathbf{F} )=&\frac{1}{2 {\kappa ( \mathbf{F} )}^{\frac{2}{\kappa ( \mathbf{F} )}-1 }} \left( \frac{2 {\kappa ( \mathbf{F} ) }^3 }{F} \right)^{\frac{2}{\kappa (\mathbf{F} )} -1}  \exp \left( -\frac{2 {\kappa ( \mathbf{F} )}^3 }{3 F} \right) \notag \\
& \times \exp \left( -6 \left(\frac{2}{{\kappa ( \mathbf{F} ) }^2} \right) \left( \frac{F}{{\kappa (\mathbf{F} ) }^3} \right) \right) ,
\label{eq:8}
\end{align}
where
\begin{equation}
\kappa ( \mathbf{F} ) = \sqrt{ 2 I_p ( \mathbf{F} ) }.
\label{eq:kappa}
\end{equation}
We note from Eqs.~\eqref{eq:2} and \eqref{eq:kappa} that $\kappa ( \mathbf{F} )$ is
a function of the strength and the orientation of the field with respect to the molecular frame.

The Keldysh parameter for the current experiment involving 
benzonitrile amounts to 1.17. It has been shown that for 
increasing molecular size the ionization becomes tunnel-like at 
larger Keldysh parameters~\cite{deWitt} ; therefore we use the tunneling 
model here. In benzonitrile the tunnel exit is far from the centre of mass, so
in circularly polarized laser fields, where no rescattering occurs, the influence of the molecular potential on the
escaping electron can be neglected. Accordingly, after the initial ionization step, the subsequent propagation of the electron can be estimated accurately by solving the classical equations of motion in the laser field only~\cite{Corkum:1993:PRL}.  We consider a left circularly polarized (LCP) probe pulse, containing more than 10 cycles, which is
switched off adiabatically at large times. These assumptions allow us to consider a field with constant amplitude
\begin{equation}
\mathbf{F} (t) = F_0 \left( \sin ( \omega t ) \hat{ \mathbf{e_y} } +
   \cos ( \omega t ) \hat{ \mathbf{e_z} } \right).
\label{eq:pulse}
\end{equation}
Here $F_0$ is the peak field strength and $\omega$  the angular frequency. The  final photoelectron momentum at the 2D detector screen can now be straightforwardly
related to the vector potential at the  instant of ionization $t_0$, that is
\begin{align}
p_Y = & - A_Y ( \omega t_0 ) = - \frac{F_0}{\omega} \cos ( \omega t_0 ) \notag  \\
p_Z = & - A_Z (\omega t_0 ) =  \frac{F_0}{\omega}   \sin ( \omega t_0 ) 
\label{eq:12}
\end{align}
where $\omega t_0$ is the angle between the Z-axis and the field vector at the instant of ionization. Using Eqs. \eqref{eq:8} and \eqref{eq:12}, for each field orientation, i.e.,  $t_0$, the final momenta in the (Y,Z) plane and the associated tunneling probability can be calculated.

\subsection{Modeling the off--the-nodal plane photoelectron emission by tunneling theory}
\label{off-nodal}

\begin{figure}
{\includegraphics[width=\figwidth]{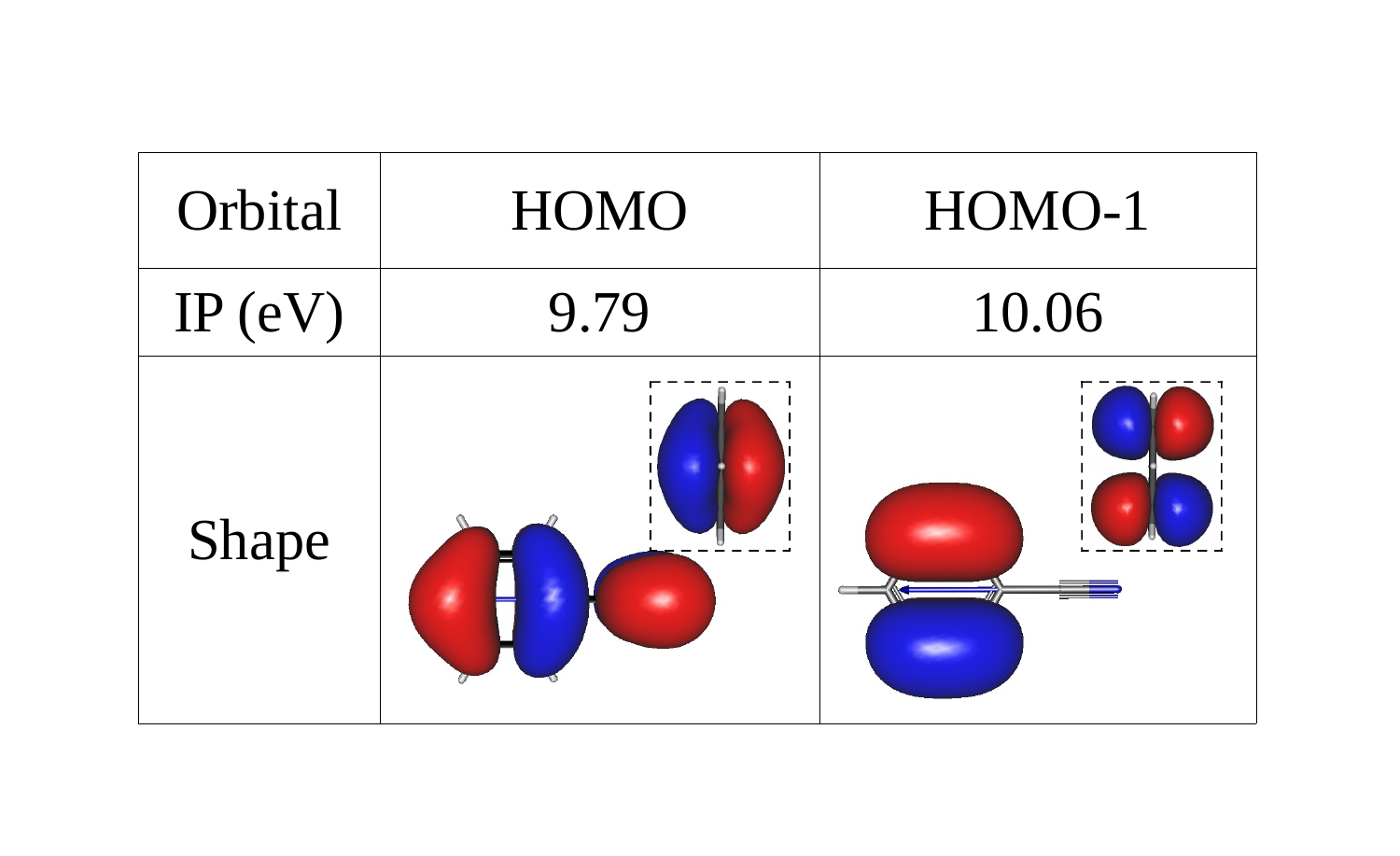}}
\caption{The HOMO and HOMO-1 of benzonitrile (C$_7$H$_5$N). The
molecular plane coincides with the (Y,Z) plane, and the detector
is on the right of these images. The insets in dashed lines, show the
view of the orbitals from the detector, showing that both orbitals
have nodal planes in the polarization plane of the circularly
polarized laser pulses. The orbitals and ionization potentials have
been obtained by quantum chemistry caculations in GAMESS~\cite{GAMESS}
at the Hartree-Fock level of theory in conjunction with the valence triple-zeta basis set.}
\label{fig:BN:orbitals}
\end{figure}

The structure of the molecular orbitals of benzonitrile with the lowest binding energies, combined with the particular 3D orientation in the present experimental setup, result in the unique X-shaped pattern in the electron momentum distribution shown in Fig. \ref{fig:BN-PADS}, and the off-the-nodal plane emission angle reported in Eq.~\eqref{omega_exp}. To explain such a pattern  requires a more advanced  treatment than the one presented in Ref. \citealp{Holmegaard:NatPhys:2010}. The HOMO and HOMO-1 of benzonitrile are shown in Fig. \ref{fig:BN:orbitals}. The most important feature for the present development is that the relevant experimental orientation of the molecule is such that the nodal planes of both the HOMO and HOMO-1 coincide with the molecular plane which in turn coincides with the polarization plane of the laser pulse. In other words, neither HOMO nor  HOMO-1  have electron density in the polarization plane of the pulse \eqref{eq:pulse}. This geometry
poses a theoretical challenge in tunneling models, since conventional tunneling and molecular tunneling theory is inapplicable. The simple
reason for their failure is that
the standard assumption, that tunneling of the negatively charged electron happens in a very narrow cone around the direction
opposite to the field~\cite{Perelomov:1966:JETP,Ammosov:1986:JETP,CDLin2002,bisgaard}, is no longer valid. In that direction
there is no bound state wave function to overlap with the continuum emerging from the outer turning point, and
no ionization will occur due to the nodal plane. In short, in the nodal plane there is no population which can tunnel, and consequently emission occur off-the-nodal plane. The simplest cases where this is evident is
tunnel emission from a hydrogenic 2p$_y$ or 2p$_x$ orbital when the field is aligned along the Z-axis.
These states are linear combinations of the states $|n=2,l=1,m=1 \rangle$ and $|n=2,l=1,m=-1
\rangle$. According to tunneling theory, the contributions from states with $m=1$ and $m=-1$ should be added
incoherently. Then, following Ref. \cite{DeloneKrainov}, the transverse momentum distribution is
proportional to
\begin{equation}
\exp \left( -\frac{ \sqrt{2 I_p  }}{F_0} p_X^2 \right) ,
\label{eq:DK_transverse}
\end{equation}
and therefore from such a state one would observe a peak at $p_X = 0$ in the momentum distributions recorded on the
detector screen in this experiment. However, as is evident from the present experiment, the
emission occurs off-the-nodal plane. Inclusion of off-the-nodal plane emission in the tunneling model is
nontrivial, since the dominant photoelectron momenta under tunneling emission (momentum distribution of tunneled electron at its birth)
should now depend on the particular initial state. For example, 2p$_y$ or 2p$_x$ orbitals would have different transverse
momentum distributions. Concerning the derivation of the tunneling rate, we can, in this situation, no longer
separate the azimuthal angle around $\mathbf{F} (t)$ and thus cannot add the individual $m$
components incoherently. We note here that even when static Stark shifts are included, the
information about the nodal plane of the initial state survives and cannot be washed out by the
polarization effects.

Motivated by earlier work on tunneling theory for large systems
\cite{Brabec:2005:prl}, and by the fact that the saddle-point formulation of the length-gauge strong-field approximation (SFA)~\cite{Keldysh}  in
the limit of  small Keldysh parameter agrees very well with the tunneling theory
\cite{Gribakin} (at least for a zero-range potential), we modify  the transverse momentum
distribution of Eq.~\eqref{eq:DK_transverse} by multiplying it by the squared Fourier transform of the
initial wavefunction. More precisely, for the 2p$_x$ orbital  that means multiplication of the transverse
momentum distribution~\eqref{eq:DK_transverse} by the factor $ ( p_{\rho} \cos (\phi) )^2 $, \ie,
\begin{equation}
w( p_{\rho} , \phi ) \sim  p_{\rho}^2 \cos^2 ( \phi )  \exp \left( - \frac{ \sqrt{2 I_p (0)}}{F_0} p^2_{\rho} \right) ,
\label{eq:18}
\end{equation}
where $p_{\rho}$ and $\phi $ are the 2-dimensional polar coordinates in the (X,Y) plane. In Eq.~\eqref{eq:18}  $ ( p_{\rho} \cos (\phi) )^2 $ is the dominating factor relating to the angular behavior of the squared Fourier
transform of 2p$_x$ at $p_Z = 0$, and a factor  $((p/\kappa)^2+1)^{-6}$ has been neglected. Finally,
in the above equation for the transverse momentum distribution, we have used the field-free ionization potential (appearing as a square root
in the exponential) since, as we have checked,  adding Stark shifts to this part of the distribution does not change the results much.

Using the distribution \eqref{eq:18} the expectation value of the X-component of the transversal
momentum (we assume the field is oriented along the Z-axis), $p_X = p_{\rho} \cos ( \phi )$ in the
interval $p_X >0 $ can be calculated  analytically as
\begin{equation}
\langle p_X \rangle = \frac{\int_0^{\infty} d p_{\rho} p^2_{\rho} \int_{-\pi/2}^{\pi /2} d \phi \cos ( \phi ) w ( p_{\rho} , \phi ) }{\int_0^{\infty} d p_{\rho} p_{\rho} \int_{-\pi/2}^{\pi /2} d \phi w ( p_{\rho} , \phi ) } = 2 \sqrt{ \frac{F_0 }{ \pi \kappa } },
\label{eq:19}
\end{equation}
where $\kappa = \sqrt{2 I_p (0)} $. We note that the expectation
value of $p_Y$ for the tunneled electron at birth is $\langle p_Y
\rangle = 0$ and that the probability of emission with $p_X = 0$ is
zero, \ie, there is no emission in the nodal plane of the initial
state although the field is along the Z-axis. The final momenta
of the tunneled electron along the Y- and Z-axis are defined
by Eq. \eqref{eq:12} and it follows that  the maximal final electron momentum along
the Y-direction is $p_Y^{max} = F_0 / \omega $. We can
define an angle $\Omega= \arctan ( \langle p_X \rangle / p_Y^{max}
)$ as a measure for the off-the-nodal plane emission. With the above expressions, this angle can be calculated analytically and the results reads
\begin{align}
\Omega &= \arctan ( \langle p_X \rangle / p_Y^{max} ) = \arctan
\left( \frac{2 \omega }{\sqrt{ \pi \kappa F_0 }}  \right) \notag \\
&= \arctan \left( \frac{2 }{\sqrt{ \pi }} \gamma
\sqrt{\frac{F_0}{{\kappa}^3}}  \right), \label{eq:20}
\end{align}
where $\gamma = \omega \kappa / F_0$ is the Keldysh parameter~\cite{Keldysh}. The above expression is,
to our knowledge, the first quantification of the angle of off-the-nodal plane emission as a function of
the field intensity and frequency and the binding energy of the atomic/molecular system. It has the
intuitively correct behavior with respect to increase of the field intensity and the binding energy:
the angle decreases. Since both peak field $F_0$ and $\kappa$ appear as square roots in the
expression, in the limit of large intensities, where also the tunnel model is more  valid, the angle is
inversely proportional to the fourth root of intensity, \ie, its dependence on intensity is very
weak.

\begin{figure}
\begin{center}
\includegraphics[width=\figwidth]{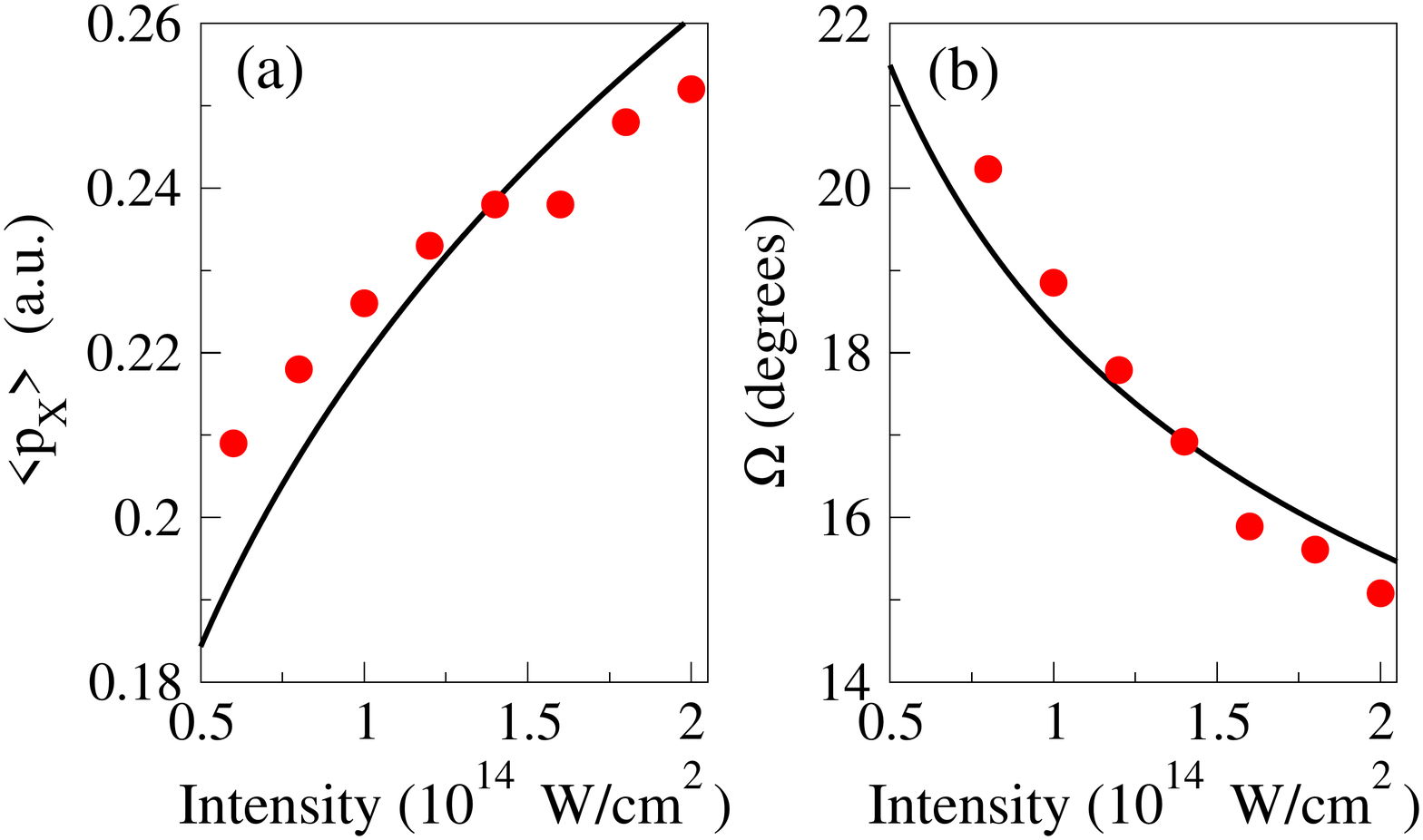}
\caption{
Expectation value of (a) $p_X$ and (b)  off-nodal-plane emission angle $\Omega$ as a
   function of intensity of the circularly polarized field for a p$_x$ orbital  with binding energy of
   $0.5$ a.u. similar to the HOMO of benzonitrile and a laser wavelength of $800$ nm. The full curve denotes the present analytic theory,
   and the dots denote the results obtained from calculation within the strong-field approximation~\cite{Keldysh}.    } \label{fig:5}
\end{center}
\end{figure}

To test our model, we imagine a fictitious experiment where we
have an 'oriented' atom in a p$_x$ orbital and the field-free ionization potential $I_p
(0) =0.5$ a.u.. We assume that, identically to the experiment
involving benzonitrile, ionization is induced  by a LCP field in the
(Y,Z) plane and that the momentum distribution is recorded on a
detector coinciding with the (X,Y) plane. The final momentum
distribution recorded on the screen can be calculated for this
case using the above off-the-nodal-plane tunnel emission model. We
have compared the results from our model to the calculations
within the strong-field approximation~\cite{Keldysh} assuming a 10 cycle circularly polarized laser pulse
with a sine-square envelope, defined in terms of the vector
potential $\mathbf{A} (t)$ interacting with the same 2p$_x$ state. The comparison between the predictions of $\langle p_X \rangle$ and $\Omega$ by the two methods is given in Fig.
\ref{fig:5}. It can be seen that for large intensities, the
tunneling model and the SFA calculations largely agree for both quantities.
As the intensity of the field decreases, or equivalently, the
Keldysh parameter $\gamma$ increases, the discrepancy between the
SFA and the tunneling model results increases, reflecting the
gradual departure from the tunneling regime.

\section{Strong-field ionization of 3D-oriented benzonitrile}

Armed with our generalized tunneling model including both the static Stark shifts and the
off-the-nodal-plane emission, we now address the experimental results for benzonitrile.

As illustrated in Fig.~\ref{fig:BN-PADS}, two sets of experiments have been performed involving pre-oriented benzonitrile molecules: one with 1D-oriented molecules and the other with 3D-oriented molecules. The results obtained with 1D and 3D orientation show almost
the same degree of up-down asymmetry (Fig. \ref{fig:BN-PADS}). The asymmetry can be ascribed to the dependence of the ionization
potential on the static Stark shifts of the energy levels (see Sec.\ref{Stark}), i.e., the small difference between the dipole moment of benzonitrile and its cation (Appendix A). The case of 1D-oriented benzonitrile can be treated theoretically by the same method as introduced in Ref.~\citealp{Holmegaard:NatPhys:2010} for the description of the linear OCS molecule. In the 1D oriented case, the molecules are allowed to rotate freely around its molecular axis and the effect of
off-the-nodal-plane emission is washed out.
In the experiment on 3D aligned and oriented benzonitrile, on the other hand, the polarization plane of the
circularly polarized laser lies entirely in the nodal plane of both HOMO and HOMO-1.
Indeed, in this case, off-the-nodal-plane emission occurs and manifests itself as an X-like pattern in the photoelectron momentum distribution on the detector screen, and
\begin{figure}
   \centering
\includegraphics[width=\figwidth]{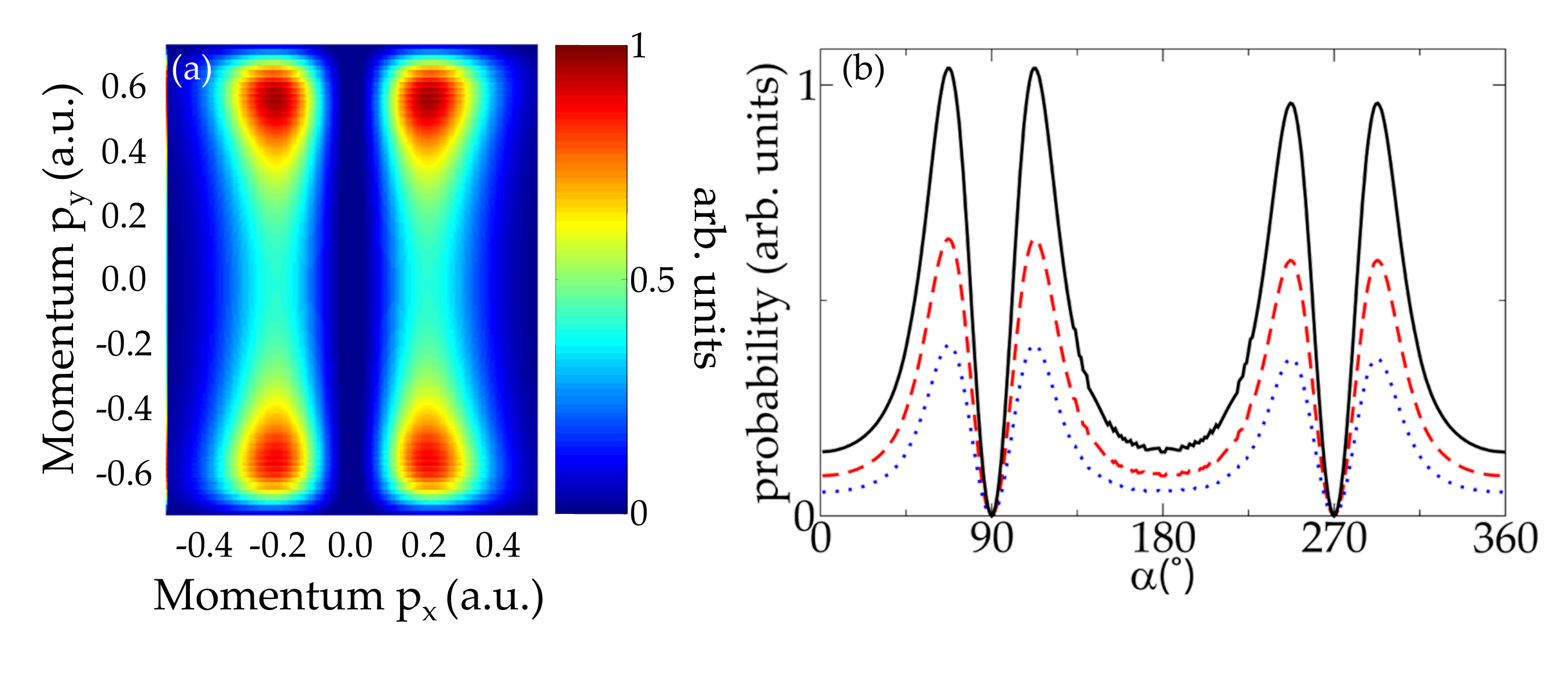}
\caption{(a) Momentum distributions including volume effects for benzonitrile at $1.23\times 10^{14}$
   W/$\rm{cm}^2$ for ionization from the HOMO and HOMO-1. (b) Angular distributions derived from momentum
   distributions. Black (full): HOMO+HOMO-1, red (dashed): HOMO, blue (dotted): HOMO-1. The results should be compared to the corresponding experimental findings in Fig. \ref{fig:BN-PADS}.
   } \label{fig:6}
\end{figure}
here we focus on the 3D-oriented benzonitrile. To describe the experiment, we use the theory for
off-the-nodal plane emission, developed in the previous section. The polarizability of
benzonitrile is very large (see Apendix A), so we expect that the initial orbital would be modified by the polarization
response of the system, and we assume that any structure of the orbital off the nodal plane will be washed out. Thus, when the Stark shift is not included, the total probability of
tunnel emission will not depend on the (finite) angle of the electric field vector with respect to the
Z-axis. Therefore, we model all orbitals of benzonitrile that have their nodal plane coinciding with
the polarization plane with the simplest possible orbital having such a nodal plane: the p$_x$
orbital. All theoretical results presented here assume perfect 3D orientation of the
molecule. Using Eqs.~\eqref{eq:8}, \eqref{eq:12} and \eqref{eq:18}, the momentum distributions
at a particular intensity can be calculated. We perform focal volume integration by including intensities
ranging from $1.2 \times10^{14}$~W/cm$^2$ to $0.5 \times10^{14}$~W/cm$^2$ with a step of $5 \times 10^{12}$
$ {\rm W/cm^2}$. The resulting momentum distributions from HOMO, HOMO-1 and the combined response of HOMO and
HOMO-1 are shown in Fig. \ref{fig:6}.


As seen by comparison of Fig. \ref{fig:BN-PADS} and Fig. \ref{fig:6}, at larger radial final momenta $p_{\rho} = \sqrt{ p^2_X + p^2_Y}$,
there is good agreement between the experimental and theoretical differential momentum distributions. Also, the agreement between
theory and experiment for the angle off-the-nodal-plane emission
$\Omega$ is good: the experiment yields $\sim 18$ degrees (Fig.
\ref{fig:BN-PADS}, [Eq.~\eqref{omega_exp}]) and the theory $18.8$ degrees [Eq. \eqref{eq:20}].
In agreement with the calculation the suppression of electron emission occurs at $\alpha$~= 90 and 270 degrees. The degree of suppression seen
experimentally is, however, much less pronounced that the complete suppression, i.e. no electron emission with $p_X = 0$, predicted theoretically.
One factor contributing to this disagreement is the lack of perfect alignment of the molecular plane in the experiment.
Another source that can produce a signal at $p_X=0 $ is
emission from the HOMO-2 that is bound with additional 0.1 a.u.
compared to the HOMO. The HOMO-2 has population in the polarization plane of the
field so that its contribution can still be significant (see Fig. 4  in the Supplementary Information of Ref. \cite{Holmegaard:NatPhys:2010}).

The present theory  is capable of describing the small up/total asymmetry observed in
the experiment by attributing it to the Stark shift of the ionization potential due to the
dipole and polarizability of the HOMO orbital, i.e., the
difference of the permanent dipole moment of the molecule
with respect to the permanent dipole moment of the unrelaxed cation. The calculated up/total asymmetry for the case of
3D-oriented molecule at $1.23\times 10^{14}$ W/$\rm{cm}^2$ is 0.52 in fair agreement with the experimental results.


\section{Conclusions}

We have studied molecular frame photoelectron distributions (MFPADs) following strong field single ionization, with circularly polarized, nonresonant fs pulses, of benzonitrile molecules aligned and oriented in space by mixed laser and static electric fields. The most striking experimental observation is suppression of electron emission in directions that correspond to nodal planes of the orbitals from which ionization occurs. The experimental findings are explained using a modified tunneling theory. Our theoretical analysis shows that the circularly polarized ionizing field drives the photoelectrons away from the cation, minimizing effects of post-ionization interaction, and leads to a clear signature of the orbital structure of the molecule at the time of ionization. In particular, we have shown here that nodal structures can be mapped out, a possibility recently examined further theoretically~\cite{Martiny2010}.

An important future challenge is to extend the present work to the study of time-dependent phenomena. This may be realized, for instance, by a pump-probe scheme where the probe pulse promotes molecules from the electronic ground state to an electronically excited state. One class of such photoinduced reactions that could be well suited for the strong field probing, introduced here, concerns large amplitude vibrational motion in the excited state. This occurs for, instance, in axially chiral systems such as biphenyl or phenyl-pyrrole \cite{Okuyama:jcp:1998} where the dynamics in the excited states corresponds to torsional motion of two aromatic rings. Alignment as well as orientation of such systems is feasible ~\cite{Madsen2009PRL,Madsen2009JCP} and the present probe method could map out the changing valence electronic structure during torsional motion occurs. More generally, strong field ionization in combination with MFPADs may provide an informative probe of charge migration and charge transfer on time scales down to a few femtoseconds \cite{Kuleff:CP:2007}.

\section{Acknowledgement}
\label{acknowledgement}

The work was supported by the Danish National Research Foundation, the Lundbeck Foundation, the
Carlsberg Foundation, the Danish Council for Independent Research (Natural Sciences) and the European Marie Curie Initial Training Network Grant No. CA-ITN-214962-FASTQUAST.

\section*{APPENDIX A}
\label{appendix}

The molecular structure, population analysis, and orbital energies of benzonitrile were obtained
from Hartree-Fock calculations~\cite{GAMESS} (the molecule is oriented in the ($Y,Z$) plane and the CN-end
pointing in the positive Z axis, see Fig. \ref{fig:BN-PADS}). The HOMO and HOMO-1 of benzonitrile are shown in Fig.~\ref{fig:BN:orbitals}. The orbital energies, given in Fig.~\ref{fig:BN:orbitals}, agree well with the experimental
ionization potentials, reported in Ref.~\cite{Kobayashi74}, and are very close in energy
[$I_p$: 9.79~eV ($0.36$ a.u) for HOMO; 10.06~eV ($0.37$ a.u) for HOMO-1]. Population analysis shows that while the
HOMO-1 is localized on the aromatic ring, about 30\% of the HOMO electron density is
on the CN-end, see Fig.~\ref{fig:BN:orbitals}. The experimental 3D orientation of the molecule is such that the nodal planes of both HOMO and HOMO-1 orbitals, coincide with the molecular plane, i.e., they have no electron density in the molecular plane.

For the purpose of our Stark-shift tunneling model we require the permanent dipole moment and the polarizabilities of benzonitrile and its cation. The nonzero  elements of the polarizability tensor of benzonitrile are (in a.u.):
${\alpha}_{xx}^{MOL}=50.522$, ${\alpha}_{yy}^{MOL}=87.808$, ${\alpha}_{zz}^{MOL}=125.758$,
${\alpha}_{xx}^{ION}=42.397$, ${\alpha}_{yy}^{MOL}=76.282$, and ${\alpha}_{zz}^{ION}=151.467$, and the
permanent dipole moment of benzonitrile is $1.791$ a.u. (in fair agreement with the experimental value \cite{Wohlfart:jms:2008}) and $1.822$ a.u. for the cation, both pointing along  $-Z$  from the detector to the repeller.

\end{document}